\documentclass[12pt]{iopart}

\usepackage{amsfonts}
\usepackage[numbers]{natbib}

\usepackage{iopams} 
\usepackage{graphicx}
\usepackage{subfig}
\usepackage{siunitx}

\begin{document}

\title[An a posteriori data-driven method for phase-averaged optical measurements]{An a posteriori data-driven method for phase-averaged optical measurements} 

\author{E. Amico$^1$, S. Montagner$^1$, J. Serpieri$^1$, G. Cafiero$^1$}

\address{$^1$ Department of Mechanical and Aerospace Engineering, Politecnico di Torino, Turin, Italy}
\ead{enrico.amico@polito.it}
\vspace{10pt}
\begin{indented}
\item[]
\end{indented}

\begin{abstract}
Phase-averaging is a fundamental approach for investigating periodic and non-stationary phenomena. In fluid dynamics, these can be generated by rotating blades such as propellers/turbines or by pulsed jets. Traditional phase-averaging approaches often rely on synchronized data acquisition systems, which might require high-speed cameras, light sources, and precise delay generators and encoders, making them expensive and sometimes unfeasible. This work proposes an \emph{a posteriori} data-driven approach that reconstructs phase information from randomly acquired uncorrelated photographic frames (snapshots) using the ISOMAP algorithm. The technique enables accurate reordering of snapshots in the phase space and subsequent computation of the phase-averaged flow field without the need for synchronization. 
The framework was validated through numerical simulations and experimental fluid dynamics datasets from an optical setup featuring single- and multi-propeller configurations. The results demonstrate that the proposed method effectively captures the periodic flow characteristics while addressing the challenges related to synchronization and hardware limitations. Furthermore, the ability to apply this technique to archival datasets extends its applicability to a wide range of experimental fluid dynamics studies.  
This approach provides a scalable and cost-effective alternative to traditional methods for the analysis of periodic phenomena.
\end{abstract}

%
%
%
%
%

\section{Introduction}

Investigating non-stationary and periodic phenomena presents notable challenges due to their different nature: While some phenomena are directly correlated with the underlying periodic mechanisms at play, these might cause/influence a cascade of effects with a much broader spectral content. An example of this occurs in fluid dynamic, especially in contexts like the wake generated by propellers/turbines or flows generated by a periodic actuation, for instance, zero-net-mass-flux jets. Such flows exhibit periodic and organized patterns, making their accurate analysis reliant on specialized techniques. One widely used method for examining these flows involves performing phase-averaged measurements/simulations, which focus on isolating the periodic characteristics by aligning data acquisition with the flow's inherent periodicity.\\
Phase-averaging techniques have been widely employed to investigate flow fields in such contexts. For example, \citeauthor{Greco_Cardone_Soria_2017} employed a methodology for phase averaging by synchronizing a laser with the piston motion in impinging jet experiments. This approach utilized high-speed lasers and cameras, ensuring precise timing and detailed capture of flow structures at various phases of the cycle. Synchronization enabled the acquisition of phase-resolved data, which is critical to understanding the periodic behavior of the flow. Similarly, \citeauthor{CAFIERO2019302} studied the evolution of the starting vortex generated by a synthetic jet when interacting with a square fractal grid. Examples of applications of phase-averaged measurements can be found in the fields of combustion \cite{TEMME2014958}, the evolution of propeller wakes \cite{Lee2004, Ragni2012} or of boundary layer instability mechanisms \cite{Serpieri2018}.
In the case of propeller-induced flows, phase-averaged flowfield studies are particularly valuable for analyzing the periodic and organized variations in the flow surrounding the blades. Such analyses are pivotal for applications in distributed propulsion systems and drones for urban air mobility (UAM), where propeller interactions significantly influence efficiency and noise characteristics. 
However, conducting phase-averaged measurements in experimental setups presents several challenges. In the case of propellers, for instance, maintaining a stable rotational speed can be particularly cumbersome, as it requires high-quality motors and sophisticated controllers, which can be difficult to implement, especially with the smaller motors used for small drones' propellers. Alternatively, capturing time-resolved flowfield data often involves the deployment of high-speed velocimetry setups with expensive lasers, cameras, and synchronizing systems.
\\To address these challenges, this article proposes an alternative method for phase-averaged analysis that does not rely on synchronized data acquisition. Instead, it utilizes an \emph{a posteriori} phase reconstruction from randomly acquired, uncorrelated data. This approach is particularly advantageous when advanced equipment is not available. Additionally, the method can be applied to previously collected datasets, extending its utility to archival data. By isolating phase information after data acquisition, this methodology offers a flexible and cost-effective solution to study periodic phenomena, expanding the accessibility to phase-averaged techniques in experimental measurements.

\section{Methodology}
\subsection{Phase averaging of optical measurements}
A $n$-dimensional manifold is a topological space in which every point has a neighborhood homeomorphic to an Euclidean space $\mathbb{R}^n$. The proposed approach employs the ISOMAP  algorithm \cite{Tenenbaum_2020} to discern a low-dimensional embedding. 

One notable application is in mass cytometry data analysis, where ISOMAP improves the separation of different cell types, leading to enhanced classification accuracy \cite{cheng2023}. In hydrogeology, it has been applied for high-resolution characterization of groundwater systems, using well log data to optimize regression parameters \cite{mohammed2025}. 

In industrial diagnostics, ISOMAP, in combination with machine learning techniques, has been successfully employed for fault diagnosis in high-speed train air compressors, improving anomaly detection accuracy \cite{peng2024}. Furthermore, in the field of brain-computer interfaces (BCI), a modified version of ISOMAP, known as Supervised Discriminant ISOMAP Projection (SD-ISOMAP), has been utilized to enhance EEG signal classification, making BCI systems more efficient and practical for real-time applications \cite{reddy2024}. 

Additionally, ISOMAP has found applications in digital marketing, where it has been used in conjunction with big data technologies to improve sentiment analysis and emotion-based consumer behavior studies on e-commerce platforms \cite{zuo2024}. 

ISOMAP has also been successfully implemented in manifold learning \cite{Farzamnik_2023}.These examples highlight the versatility of ISOMAP in diverse domains.

\begin{figure}[htbp]
    \centering
    \subfloat[][]
    {\includegraphics[width=.5\columnwidth]{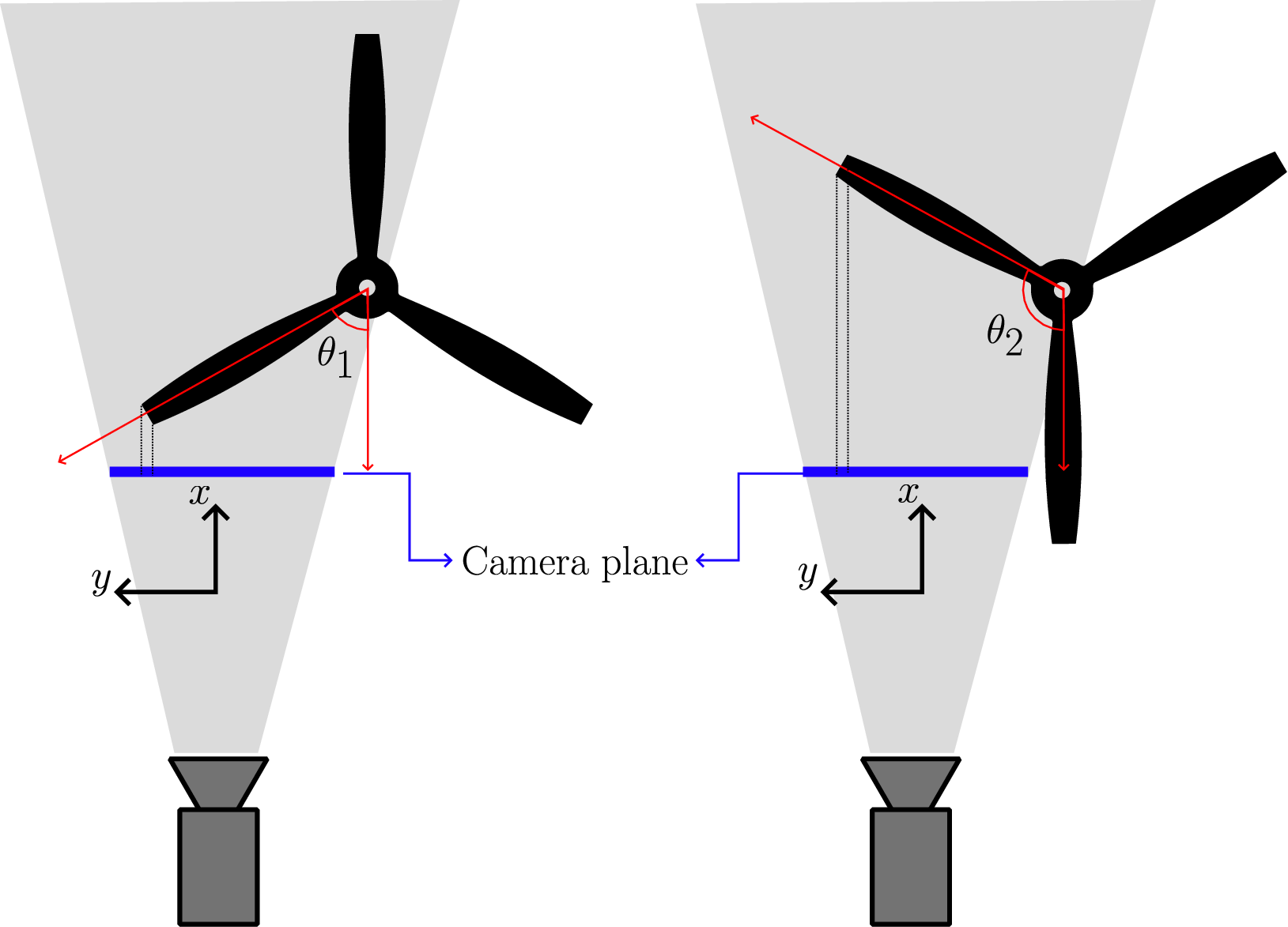}}
    	\caption{Schematic representation of the phase non-uniqueness problem observed from a plane normal to the plane of the propeller. The grey area represents the camera's viewing angles. The dashed lines represent the projection of the blade tip onto the camera plane.}
	\label{fig:prop_schema}
\end{figure}

\begin{figure}[htbp]
    \centering
    \subfloat[][]
    {\includegraphics[width=\columnwidth]{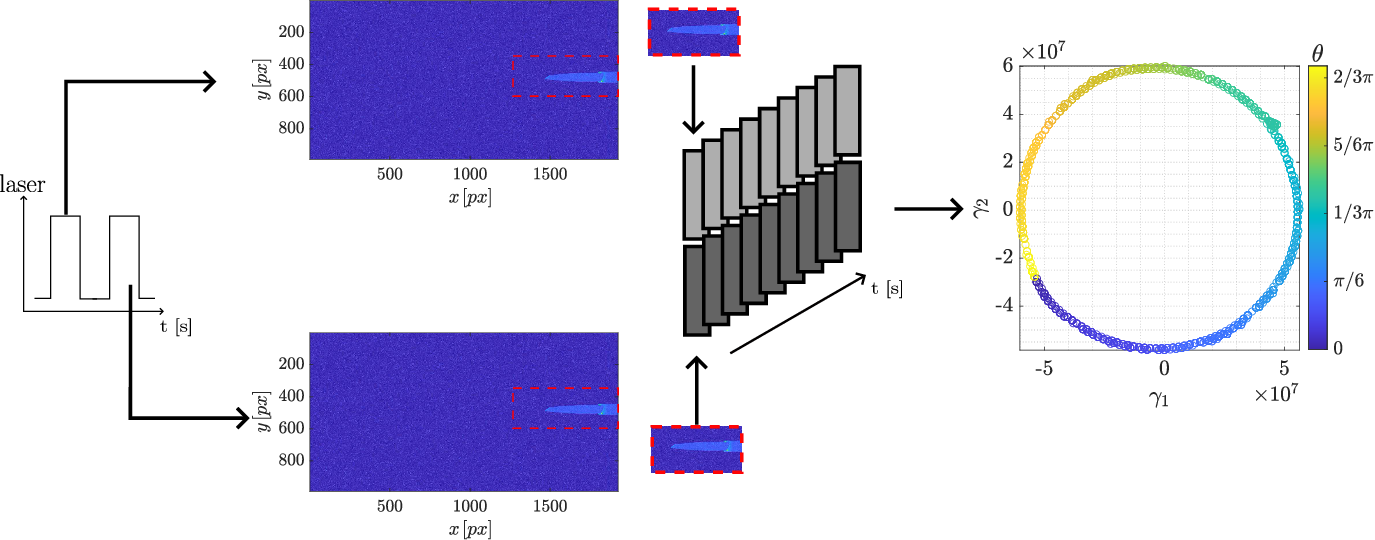}}
    	\caption{Schematic representation of the framework developed and employed for the phase reconstruction starting from the data.}
	\label{fig:framework}
\end{figure}
Particle image velocimetry (PIV) involves acquiring for each vector field a pair of particles' images with a known time delay, $\Delta_T$. The value of $\Delta_T$ is selected to ensure particle correlation, generally leading to a displacement between the two snapshots of about 10-12 pixels \citep{Willert1991, Raffel2018}. Stereoscopic PIV (S-PIV) requires two independent views of the region of interest, thus allowing the measurement of the out-of-plane component of the velocity field \citep{Willert_1997} through geometric corrections. Therefore, in the S-PIV case, each vector field requires two pairs of snapshots, one pair per camera. 

The purpose of the present study is to investigate the wake generated by a propeller operating under different inflow conditions. Since the study is primarily focused on the flowfield in the immediate vicinity of the blade, the propeller is imaged within the field of view (FOV). The plane considered is perpendicular to the propeller's rotation plane and intersects its axis of rotation.

Considering the expected velocities within the field of view \cite{grava2024experimental} and the optical setup features, $\Delta_T$ is on the order of tens of $\mu \, \mathrm{s} $; on the other hand, the characteristic time of the propeller, $T_p$, i.e. the time required to complete one revolution, is on the order of $10^{-3} \, \mathrm{s}$, considering rotational speeds of the propeller ranging from 1000 to 5000 $rpm$. Consequently, two snapshots belonging to the same pair contain information about the propeller’s rotational motion. Therefore, the particles' images include both the information about the propeller’s azimuthal position $\theta$ and the direction of rotation.  

For the S-PIV snapshots, if they both capture the propeller, the same phase will have a double representation for the same time instant. This additional information allows for a unique determination of the blade position. 
Instead, the problem persists if PIV snapshots are considered. When observing the blade from a plane normal to the propeller's plane, the positions $\theta = \theta_n$ and $\theta = 180^\circ - \theta_n$ are indistinguishable, as depicted in Figure \ref{fig:prop_schema}. This ambiguity can be resolved by considering both PIV snapshots, i.e., both the snapshots corresponding to the first and second laser pulses. In fact, if we consider the blade rotating in a clockwise direction, as shown by the red-arrowed blade in the left cartoon of Figure \ref{fig:prop_schema}, the tip moves towards higher $y$ values when transitioning from the first snapshot to the second (i.e. within very small time delays). Conversely, in the right cartoon of Figure \ref{fig:prop_schema}, the tip moves towards lower $y$ values between the two proximate snapshots.
By exploiting this information contained in the snapshots, it is possible to reorder the snapshots in ascending order in the phase space such to perform phase averaging of the reordered data. This is schematically represented in Figure \ref{fig:framework}.

\subsection{ISOMAP} \label{isomap}

\subsection*{Introduction}

The snapshots are here treated as vector functions $\mathbf{s}(\mathbf{x}) = (s_{t}(x,y), s_{t+\Delta t}(x,y))$ in a Hilbert space. The inner product between the grey-scael level of two snapshots $\mathbf{s}_i$ and $\mathbf{s}_j$ is defined as:
\begin{equation}
\langle \mathbf{s}_i, \mathbf{s}_j \rangle = \int\int_{\Omega} \mathbf{s}_i(\mathbf{x}) \cdot \mathbf{s}_j(\mathbf{x}) \mathrm{d}x \mathrm{d}y,
\end{equation}
where $\cdot$ denotes the scalar product in the two-dimensional vector space, and $\Omega$ indicates the domain.
Norms are canonically defined as $||s|| = \sqrt{\langle s, s \rangle}$, and distances between snapshots are consistent with this norm. 

To analyze a collection of $N_s$ snapshots, we employ the ISOMAP algorithm to construct a low-dimensional representation that preserves the nonlinear relationships within the dataset. The procedure consists of several key steps, as described below.

The geodesic distance matrix \(D_G \in \mathbb{R}^{N_s \times N_s}\) contains the geodesic distances between the snapshots. To compute this matrix, we first construct a \(k\)-nearest neighbors (kNN) graph, where each snapshot is connected to its \(k_e\) closest neighbors. The distances within each neighborhood correspond to Hilbert-space distances. For snapshots that are not directly connected, geodesic distances are approximated as the shortest path through the kNN graph using the Floyd-Warshall algorithm \cite{Floyd62}. 

The choice of the number of neighbors \(k_e\) is critical for building the kNN graph and approximating the geodesic distances \cite{Samko2010AutomaticCO,Farzamnik_2023,Marra_2024}. A small \(k_e\) may lead to disconnected regions, while a large \(k_e\) may cause short-circuiting, reducing the graph to an Euclidean representation and compromising ISOMAP's ability to capture nonlinear relationships. To determine \(k_e\), we monitor the Frobenius norm, as suggested by \citeauthor{Marra_2024} \cite{Marra_2024}, of the geodesic distance matrix \(D_G\), denoted as \(\|D_G\|_F\). As \(k_e\) increases, \(\|D_G\|_F\) decreases monotonically until a minimum value is reached. Beyond this point, further increases in \(k_e\) cause a sudden drop in \(\|D_G\|_F\), which suggests short-circuiting. The acceptable range for \(k_e\) is between the minimum value ensuring all nodes are connected and the value where short-circuiting occurs.

\subsection*{Construction of the low-dimensional embedding}

After obtaining the geodesic distance matrix \(D_G\), multidimensional scaling (MDS) is applied, following the method proposed by Torgerson (1952), to generate a low-dimensional embedding denoted as \(\Gamma \in \mathbb{R}^{N_s \times n}\), where \(n\) is the dimension of the embedding. The matrix \(\Gamma = (\tilde{\gamma}_{ij})\) holds the ISOMAP coordinates, with the \(j\)-th column representing the \(j\)-th ISOMAP coordinate across all \(M\) flowfield snapshots.
Each ISOMAP coordinate is represented as \(\gamma_j\), for \(j = 1, \ldots, n\). Hence, the element \(\tilde{\gamma}_{ij}\) corresponds to the value of the \(j\)-th ISOMAP coordinate for the \(i\)-th snapshot, i.e., \(\tilde{\gamma}_{ij} = \gamma_j(i)\), where \(i = 1, \ldots, M\).
This embedding captures the structure of the high-dimensional data in a lower-dimensional space, ensuring that the geodesic relationships between snapshots are preserved as accurately as possible.
Once \(D_G\) is computed, the matrix is transformed into a centered representation using:
\[
B = -\frac{1}{2} H (D_G \odot D_G) H,
\]
where \(\odot\) denotes the Hadamard product, and \(H = I_{N_s} - \frac{1}{{N_s}} 1_{N_s}\) is the centering matrix. Here, \(I_{N_s}\) is the identity matrix, and \(1_{N_s}\) is an \({N_s} \times {N_s}\) unitary matrix.
We then perform spectral decomposition of \(B\) to obtain its eigenvalues \(\Lambda\) and eigenvectors \(V\):
\[
B = V \Lambda V^\top.
\]
The \(n\) largest positive eigenvalues \(\Lambda_n\) and their corresponding eigenvectors \(V_n\) are selected. The low-dimensional embedding is calculated as:
\[
\Gamma = V_n \Lambda_n^{1/2},
\]
where \(\Lambda_n^{1/2}\) is a diagonal matrix containing the square roots of the \(n\) largest eigenvalues.

The quality of the low-dimensional representation is assessed using the \textit{residual variance}, defined as:
\begin{equation}
R_v = 1 - R^2\big({vec}(D_G), {vec}(D_E)\big),
\end{equation}

\noindent where $R^2$ is the squared correlation coefficient, ${vec}$ is the vectorization operator, and $D_E$ is the Euclidean distance matrix in the low-dimensional space. The dimensionality $n$ is chosen by identifying an elbow in the residual variance plot, which balances complexity and accuracy \cite{Marra_2024}.

The ISOMAP algorithm provides a robust method to construct low-dimensional embeddings that preserve geodesic relationships in the original data. The careful selection of \(k_e\) and \(n\) is crucial, and metrics such as the Frobenius norm and residual variance offer objective criteria to optimize the process \citep{Kouropteva2002SelectionOT}.

\subsection*{Phase sorting}
A key advantage of ISOMAP is its ability to preserve Euclidean distances between the original space and the reduced-dimensional space. If we fix a region in the image and capture multiple frames with the blade at different phases, the only variation between them will be the phase itself. As a result, images that are close together in the reduced space correspond to blade images with similar phases. Following the trajectory in the low-dimensional space naturally traces the phase progression in either a clockwise or counterclockwise direction, depending on the chosen orientation.

\section{Approach evaluation on numerical data}
\subsection{Numerical setup}
\begin{figure}[htbp]
    \centering
    \subfloat[][]
    {\includegraphics[width=.45\columnwidth]{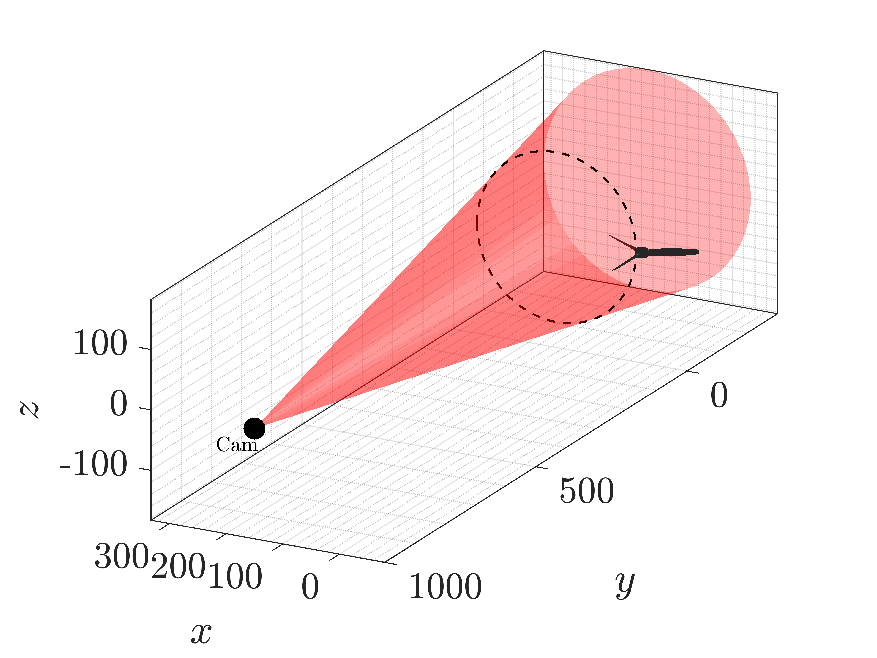}}
    \subfloat[][]
    {\includegraphics[width=.45\columnwidth]{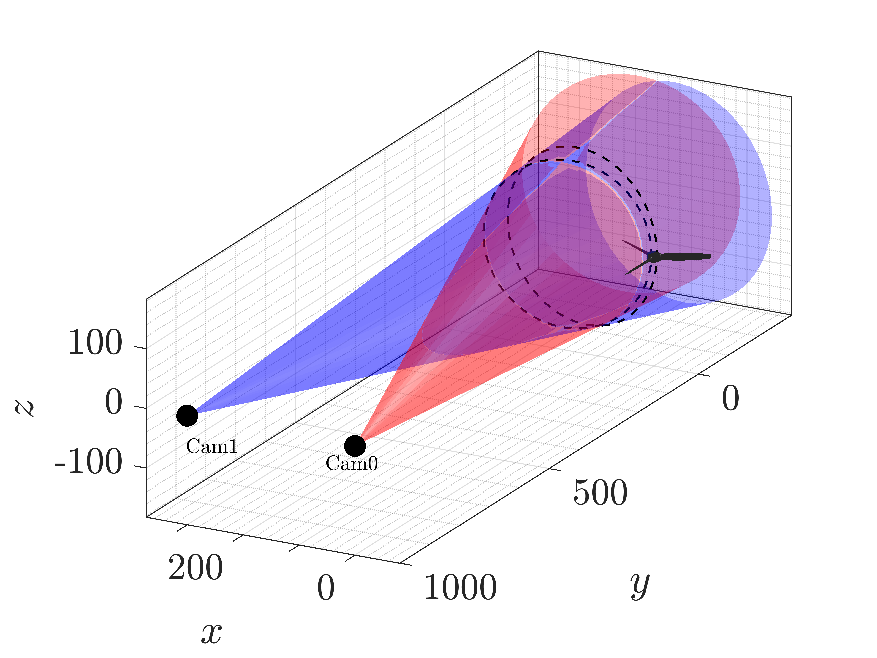}}\\
    	\caption{Representation of the numerical setup for the case of a single propeller with 3 blades, respectively PIV (a) and S-PIV (b) setups.}
	\label{fig:num_setup}
\end{figure}

To test the framework's functionality, to assess its performance, and to identify its limitations, a numerical setup was defined to replicate the experimental setup optical characteristics as accurately as possible. In particular, the performance will be evaluated for one propeller in PIV and S-PIV configuration, respectively shown in Figures \ref{fig:num_setup}a and \ref{fig:num_setup}b.

We are interested in obtaining phase-averaged values using a low-frequency optical diagnostics system. In theory, given the angular velocity of the propeller $RPM_{prop}$ ($f_{prop} =  RPM_{prop}/60$), to perform phase averaging, the PIV acquisition frequency $f_{PIV}$ must be chosen such to be a multiple of $Nf_{prop}$, where $N$ is the number of blades. 
\cite{Greco_Cardone_Soria_2017}. However, in the experiments, it is not possible to keep a constant rotational speed throughout the test duration owing to the errors introduced by the electronic speed controller (ESC) and by the motor. Despite the presence of a PID controller, in the experimental framework that will be presented below, the value of $RPM_{prop}$ could be kept constant to within $5\%$ of the target value (${RPM}_{prop}^{tv}$), which would hinder the quality of the phase averaging process. 

To account for this in the numerical setup, the following variation law for $RPM_{prop}$ is considered:

\begin{equation}\label{eq:rpm}
{RPM_{prop}}(t) = {RPM}_{prop}^{tv} \cdot \left( 1 + (\xi - 0.5) \cdot |\varepsilon_{am}| \right), \quad \xi \sim \mathcal{U}(0, 1)
\end{equation}

\noindent where $\xi$ is a random variable which can assume values between 0 and 1, while $|\varepsilon_{am}|$ is the maximum error, which is fixed to $\pm5\%$ of the target $RPM_{prop}^{tv}$ value in the present case.  From eq. \ref{eq:rpm}, it is possible to obtain  $\omega(t) = \frac{60}{2\pi \cdot {RPM_{prop}}^{tv}(t)}$. The phase $\theta(t)$ of the propeller at a given time $t$ is the time integral of ${RPM}_{prop}^{tv}(t)$, starting from an initial phase $\theta_0$, as shown in equation \ref{eq:compute_phase}.

\begin{equation} \label{eq:compute_phase}
\theta(t) = \theta_0 + \int_0^t \omega(t') \, dt'
\end{equation}

\begin{figure}[htbp]
    \centering
    \subfloat[][]
    {\includegraphics[width=.5\columnwidth]{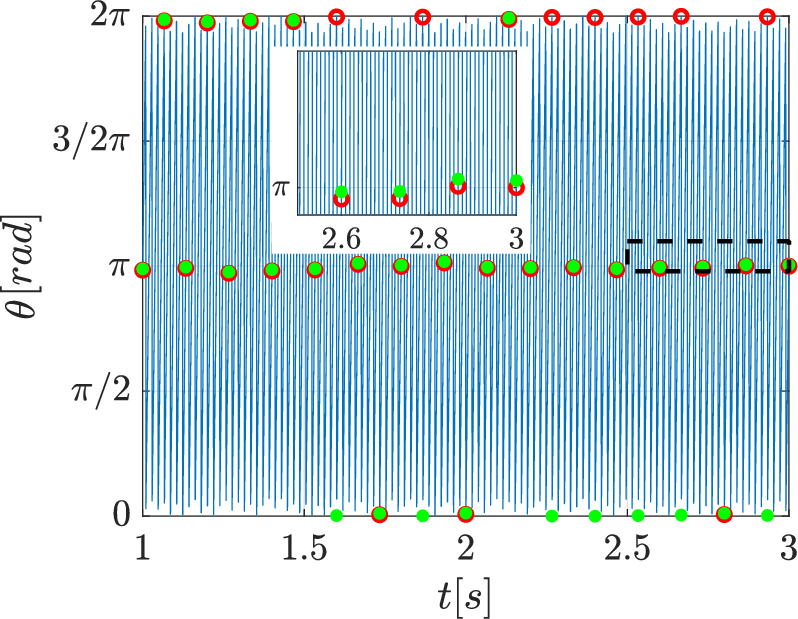}}
    	\caption{The phase progression of the propeller is shown in blue. In red and in green are indicated the time instances when images related to the first and second laser pulses are captured, respectively.}
	\label{fig:signal_trig}
\end{figure}

This approach allows for a very accurate simulation of what occurs in the experimental setup. In Figure \ref{fig:signal_trig}, the phase evolution is shown for a case with $RPM_{prop} = 1000$, $f_{PIV} = 15$ Hz, and $\Delta_t = 50\,\mu$s. The phase corresponding to the first laser pulse is shown in red, and the phase corresponding to the second pulse is shown in blue. It is possible to highlight the phase-shifting due to the $RPM_{prop}$ variation over time.

\begin{figure}[htbp]
    \centering
    \subfloat[][]
    {\includegraphics[width=.45\columnwidth]{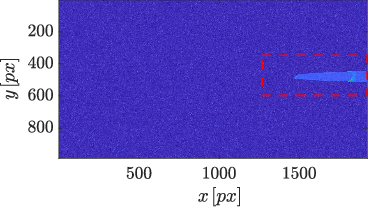}}
        
    	\caption{Example of PIV snapshots. The cropping area used to construct the matrix on which ISOMAP is applied is highlighted in red.}
	\label{fig:example_gen_img}
\end{figure}

An example of an image generated using the numerical setup is shown in Figure \ref{fig:example_gen_img}. A resolution of $1924$ [pixels] $\times$ $990$ [pixels] was chosen and a pixel depth of $16$ bits is considered. Gaussian noise with a standard deviation equal to $7\%$ of the maximum value of the image was added to simulate the effect of random noise on the gray-scale image.

\subsection{Results}
The results will be presented for the cases of one isolated propeller and two propellers. In both cases, the propellers shall have three blades, without leading the generality of the presented results. 
For the single propeller with 3 blades, 1050 image pairs were generated, corresponding to $RPM_{prop} = 3150$, $f_{PIV} = 15$ Hz, and $\Delta_t = 50\,\mu$s. Since we are interested in determining the phase of the propeller, only a portion of the entire image is considered, as indicated by the red dashed line in Figure \ref{fig:example_gen_img}.

\begin{figure}[htbp]
    \centering
    \subfloat[][]
    {\includegraphics[width=.45\columnwidth]{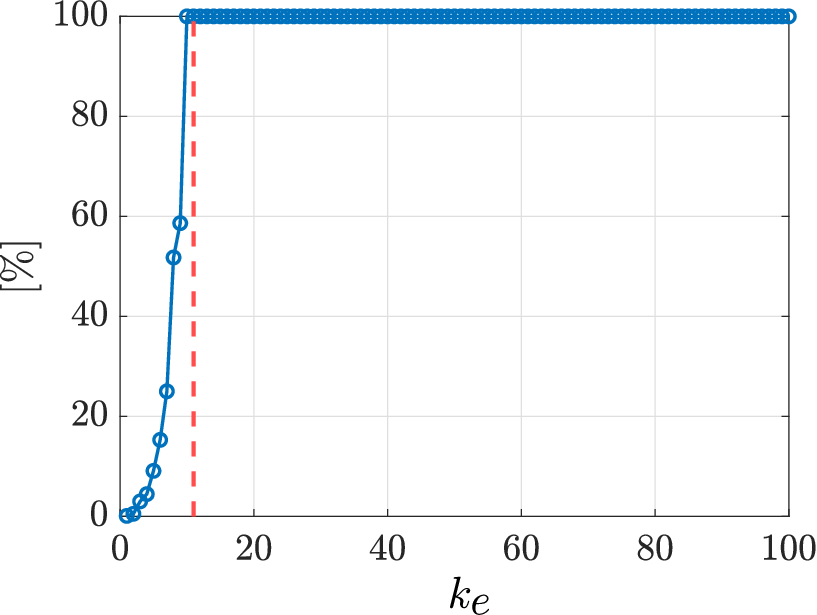}}
    \subfloat[][]
    {\includegraphics[width=.45\columnwidth]{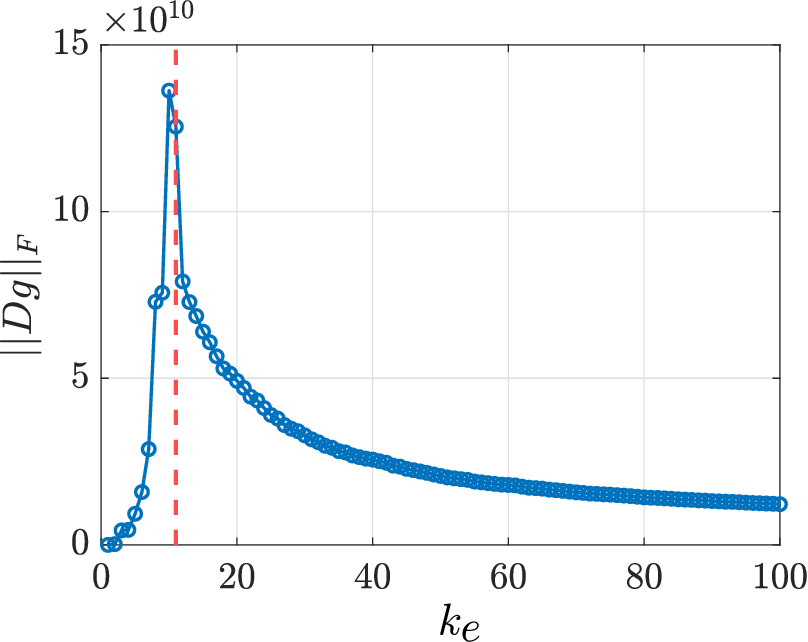}}\\
    	\caption{(a) The ratio of connected snapshots to the total number of snapshots as $ke$ increases.(b) The Frobenius norm of the geodesic distance matrix plotted as a function of the number of neighbors used in Floyd's algorithm. }
	\label{fig:norm_frob}
\end{figure}

The ratio of connected snapshots to the total number of snapshots as $k_e$ increases is shown in Figure \ref{fig:norm_frob}(a): from $k_e = 10$ the snapshots are fully connected. The Frobenius norm of the geodesic distance matrix plotted as a function of the number of neighbours used in Floyd's algorithm is reported in Figure \ref{fig:norm_frob}(b). The result shows that for values of $k_e > 11$ the short-circuiting occurs. As such, following \citeauthor{Marra_2024}, the value of $k_e = 11$ is selected.

Once the reduced-dimensional space is reconstructed, considering that new measurements are to be taken under the same conditions in terms of $RPM_{prop}$, $f_{PIV}$ (Hz), and $\Delta_T$, it is not necessary to recalculate the reduced-dimensional space. Instead, it is sufficient to project the new snapshots onto this space. 

\begin{figure}[htbp]
    \centering
    \subfloat[][]
    {\includegraphics[width=.5\columnwidth]{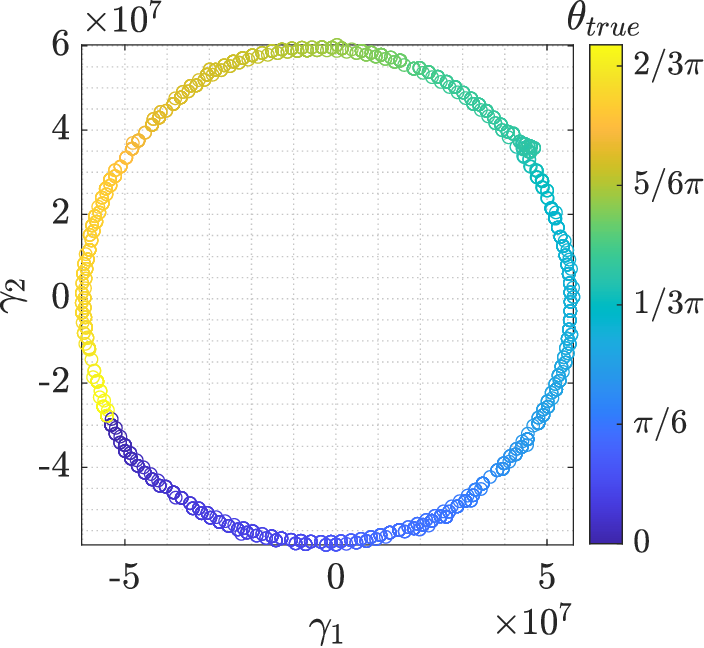}}
    \subfloat[][]
    {\includegraphics[width=.465\columnwidth]{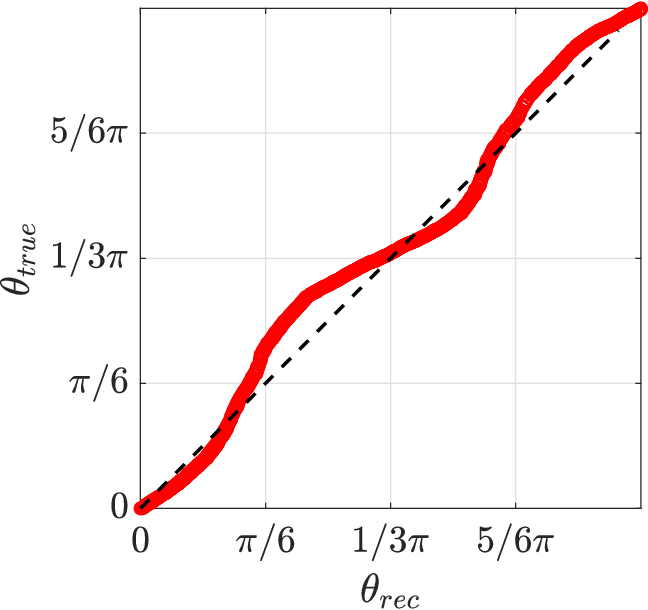}}
    	\caption{(a) Representation of the snapshots in the two-dimensional space obtained through ISOMAP, with each point labelled according to its phase. (b) $\theta_{true}$ vs. $\theta_{rec}$; the reference bisecting line, shown in red, corresponds to the error-free points loci.}
	\label{fig:iso_3blade_res}
\end{figure}

The result of the ISOMAP is shown in Figure \ref{fig:iso_3blade_res}. It is important to note that the propeller has three blades, resulting in a symmetry that reduces the possible phase values to $0 \leq \theta \leq 2\pi/3$. As illustrated in the figure, the snapshots in the reduced-dimensional space, identified by $\gamma_1$ and $\gamma_2$, are arranged on a circle. Moving along the circle corresponds to a shift in the phase, as indicated by the color-coding selected for the representation of the points of figure \ref{fig:iso_3blade_res}a. 

In Figure \ref{fig:iso_3blade_res}b, the true phase value $\theta_{true}$ is plotted as a function of the reconstructed value $\theta_{rec}$. To obtain $\theta_{rec}$, it is assumed that the points identified in the reduced-dimensional space ($\gamma_1$ and $\gamma_2$) are arranged on a circle. The center of the circle is determined, and each point is transformed into polar coordinates. These coordinates are then rescaled between $0$ and $2\pi/3$, owing to the system's symmetry. The phases obtained in this way are subsequently reordered to match the minimum value of the reconstructed angle with the minimum true angle, i.e.   $min(\theta_{true}) = min(\theta_{rec})$. From Figure \ref{fig:iso_3blade_res}b, it can be observed that the proposed framework successfully reorders the images to ensure a monotonically increasing phase.

\subsection{Phase Average}
\label{subsec:Phase_Average}
After sorting the images according to the phase $\theta$, using the ordering obtained from ISOMAP (see Figure \ref{fig:iso_3blade_res}a), it is possible to perform phase averaging. Let $N_{phase}$ be the number of phases into which the $\theta$ interval is divided. For example, in the case of a propeller with 3 blades, $\theta$ varies from $0$ to $2\pi/3$. Fixing $N_{phase}$, the phase increment is defined as $\Delta_{\theta} = \frac{2\pi}{3N}$. 

For $i = 1, \dots, N$, all phases falling within the interval $(i-1) \cdot \Delta_{\theta} \quad$ and $\quad i \cdot \Delta_{\theta}$ are identified. To compute the phase average, a weighted average of the flow field is performed using a Gaussian weighting window, centered at the midpoint of the interval, with $\sigma$ equal to half the interval length. To ensure continuity, the interval over which the Gaussian filter is applied is extended. The extended interval is given by $\big[(i-1) \cdot \Delta_{\theta} \cdot (1-\epsilon), \; i \cdot \Delta_{\theta} \cdot (1+\epsilon)\big]$, where $\epsilon$ is fixed to $5\%$, i.e., $\epsilon = 0.05$.

\section{Approach evaluation on experimental data}

\subsection{Experimental setup}
\begin{figure}[htbp]
    \centering
    \subfloat[][]
    {\includegraphics[width=.5\columnwidth]{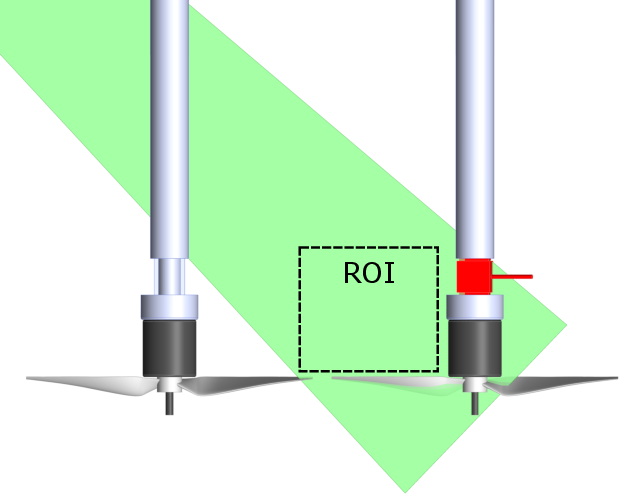}}
    \subfloat[][]
    {\includegraphics[width=.45\columnwidth]{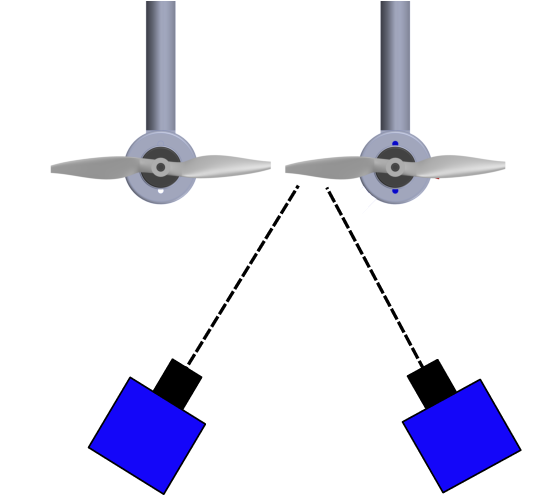}}
    	\caption{Schematic representation of the Stereo PIV experimental setup. (a) Top view. (b) Front view.}
	\label{fig:exp_setup}
\end{figure}

The experimental validation of the developed approach was obtained using a S-PIV dataset of two propellers operating in close proximity. The experiments were carried out in the Ferrari wind tunnel, an open-jet facility at Politecnico di Torino.
The reference propeller is a two-bladed rotor and its geometry was derived from the design used in \cite{CASALINO2021106707}, scaled down to a radius of $R=$ \SI{75}{\,\milli \meter} for the current application, resulting into a range of Reynolds numbers based on the chord at 75\% of $R$ varying between $1\cdot10^4$ and $3\cdot10^4$. 
The drivetrain of each propeller is powered by an RCS-TRX 370 2826 1000KV brushless motor, capable of delivering a maximum power of \SI{150}{\watt}. The motors are individually managed via external electronic speed control units, which receive command signals from a shared Arduino Uno board. The RPM of each motor is regulated by a custom PID controller implemented on the Arduino board which ensures that the average standard deviation of the measured RPM remained below 2\% of the target value.
The propellers were installed in the test section using 3D printed supports, which ensured the alignment of the rotational axes of the propellers parallel, and a minimum intrusiveness in the flowfield.\\
A schematic representation of the S-PIV setup is shown in Figure \ref{fig:exp_setup}.
The S-PIV setup was constituted by a Dantec Dynamics Nd:YAG Dual Power laser, operating at \SI{15}{\hertz} in dual pulse mode, to illuminate the plane between the two propellers. The laser provided a maximum energy of 200 mJ per pulse, ensuring adequate illumination of the tracing particles. The particles were generated using a Laskin nozzle which produced particles of 1\,$\mu$m in diameter from DEHS fluid, at a constant rate. This ensured a uniform seeding of the test section. \\Two Andor Zyla 5.5 MPx sCMOS cameras (sensor resolution: 2560 × 2160 pixels$^2$, pixel size: \SI{6.5}{\,\micro \meter}) equipped with Tokina ATX-I macro lenses (\SI{100}{\,\milli \meter} focal length) captured the light scattered by the tracing particles. Each camera was equipped with a Scheimpflug adapter. 
The region of interest (ROI) corresponds to the common region visible to both cameras, and it extends for \SI{0.9}{R} along the stream-wise direction and \SI{1}{R} in the plane containing both rotation axes. However, the images captured by each camera separately also include portions of the propeller blades. This extended field of view is crucial for the subsequent reconstruction of the propeller's phase from the captured images, as it allows the identification of blade positions relative to the flowfield. \\ The processing of the collected images was performed using the PaIRS software \cite{astarita2022pairs}. The geometric calibration was performed using a commercial double-plane target with white dots on a black background. A pin-hole method was implemented and the resulting maximum calibration error was of 0.6 pixels. The cameras calibration was then further improved by deploying a self-calibration using the light scattered by the tracing particles \cite{Wiekene_2005} \cite{Willert_1997} . This allowed us to further reduce the calibration error to 0.1 pixels. The correlation was performed using a multi-pass algorithm, with a final interrogation window size of $32\times32$ pixels$^2$ with a $75$\% overlap.

\subsection{Image pre-processing}
To better isolate the propeller blade and facilitate its position identification, the reflections produced by the blade when illuminated by the laser beam were exploited. Depending on whether the blade is illuminated or not, and on the portion of the blade that is within the illuminated plane (thus depending on the blade’s position relative to the camera plane), the resulting reflections change. These variations in reflection patterns were used as a key feature to distinguish the blades from the surrounding flowfield and to determine their phase position with high accuracy. To achieve this, the anisotropic diffusion method was applied \cite{Adatrao_2019}. This technique isolates reflections by allowing image intensity to diffuse only along edges, preserving sharp boundaries while eliminating noise and background gradients. From each original image, the method produces two distinct outputs: a background image, where unwanted reflections are captured and separated, and a cleaned image, containing only the tracer particles, with reflections effectively removed. The background image is used to identify the blades' position and accurately detect their phase. The reflections-cleaned images, on the other hand, are utilized in the S-PIV process to ensure precise measurements of the flow field.\\ By addressing the issue of unsteady reflections, the anisotropic diffusion approach enhances the reliability of both phase detection and flowfield analysis.

\subsection{Results}
\begin{figure}[htbp]
    \centering
    {\includegraphics[width=.7\columnwidth]{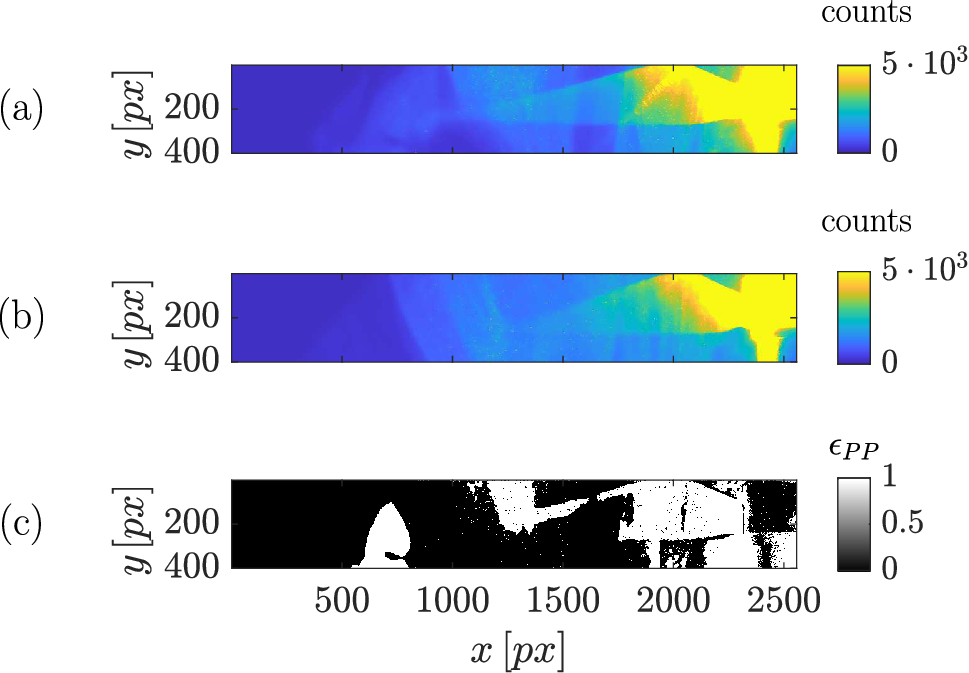}}\\

    	\caption{(a) Image of the blade region of two different PIV images selected such as (b) is separated from (a) by a finite number of propeller revolutions, based on the target RPM value. (c) Difference between the two images in terms of $\epsilon_{PP}$ (Equation  \ref{eq:epsilon}).}
	\label{fig:PIV_images_exp}
\end{figure}

\begin{figure}[htbp]
    \centering
  {\includegraphics[width=.5\columnwidth]{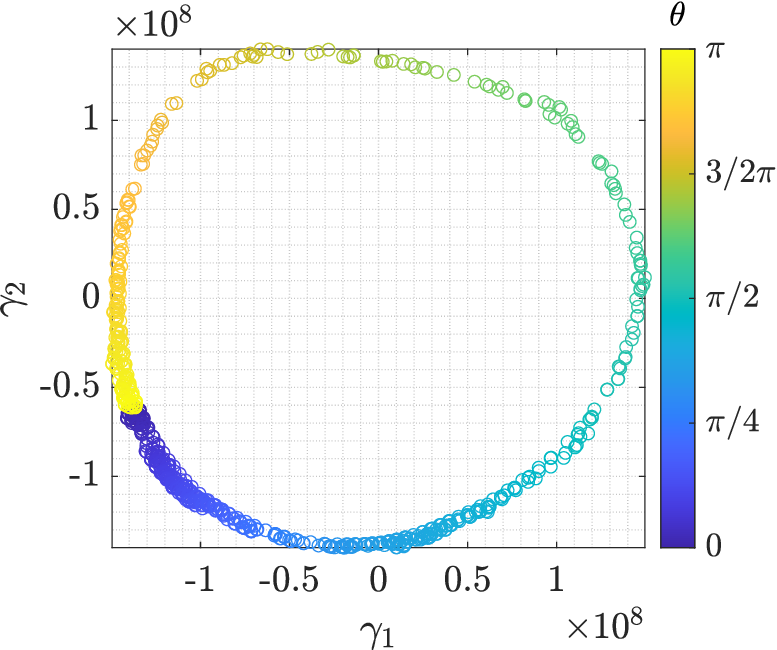}}   
    	\caption{Representation of the experimental background images in the two-dimensional space obtained through ISOMAP.}
	\label{fig:Isomap_exp}
\end{figure}

\begin{figure}[htbp]
    \centering
    \subfloat[][]
    {\includegraphics[width=.45\columnwidth]{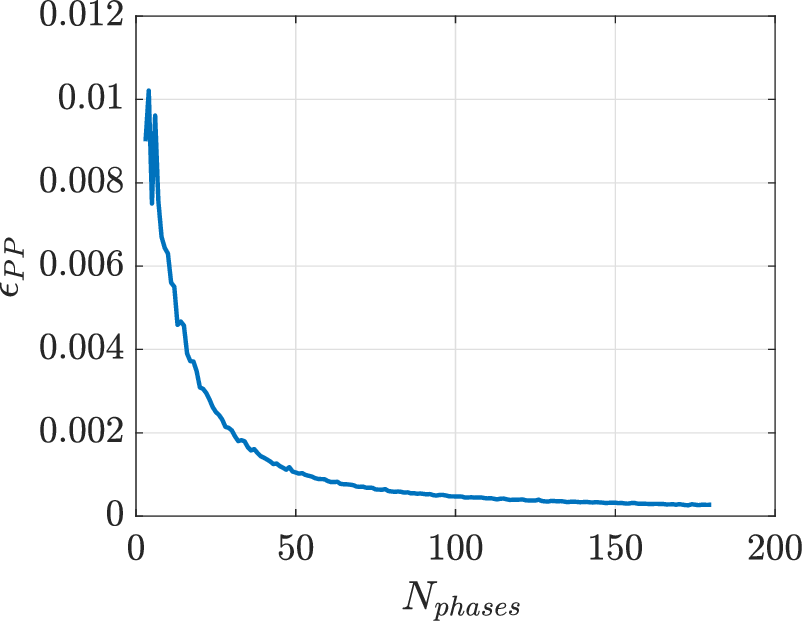}}
    \subfloat[][]
    {\includegraphics[width=.45\columnwidth]{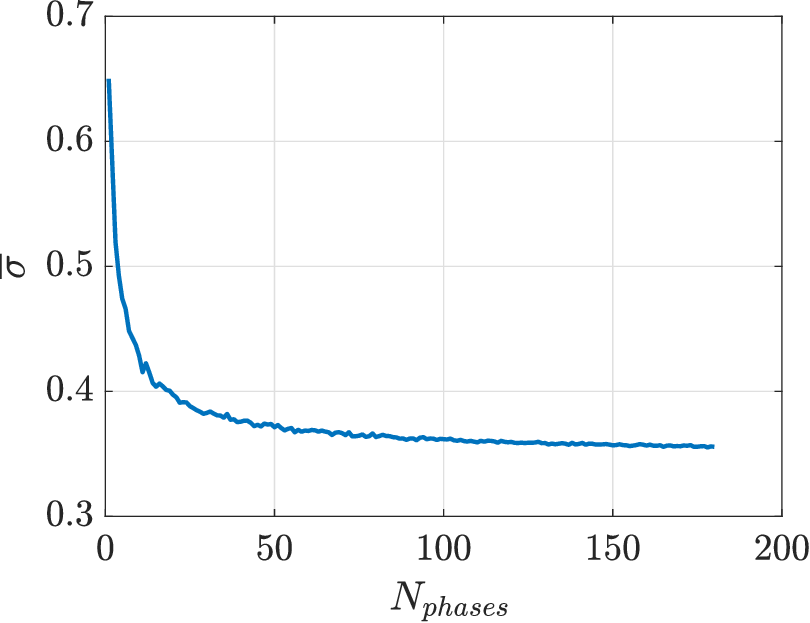}}
    	\caption{Influence of the number of phase clusters $N_{phase}$ on the phase-averaging process analyzed by evaluating the error $\epsilon_{PP}$ (a) and the standard deviation (b).}
	\label{fig:error}
\end{figure}

\begin{figure}[htbp]
    \centering
  {\includegraphics[width=0.9\columnwidth]{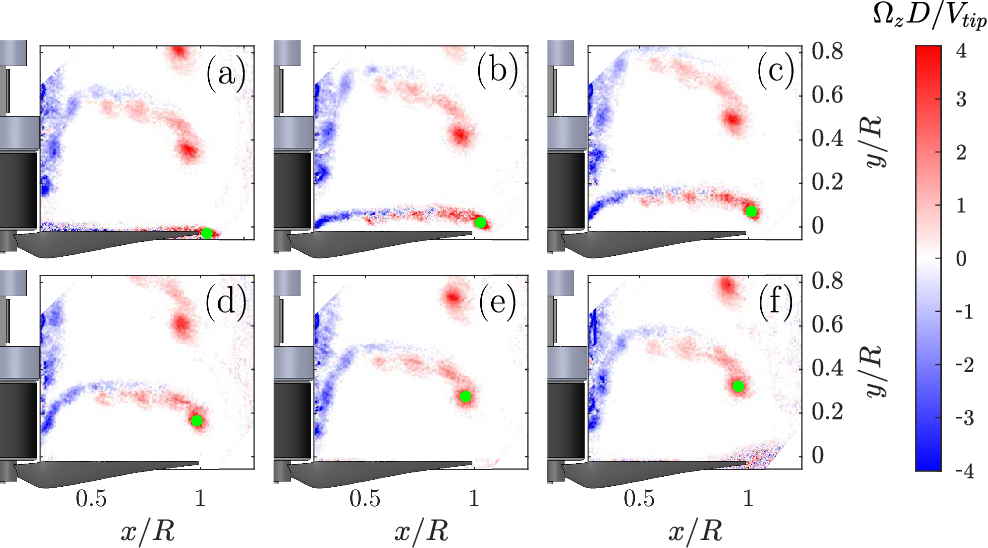}}   
    	\caption{Out of plane vorticity $\Omega_z$ normalized with respect to $V_{tip}/D$, for six different ordered phases.}
	\label{fig:omega}
\end{figure}

The results presented below refer to the case of a single propeller with a target RPM set to \SI{6000} and operating at an advance ratio of $J=0.3$, while the PIV acquisition frequency is $f_{PIV}=$\SI{15}{\hertz}. 
Two different PIV images, properly cut to highlight the blade region,   
are shown in Figure \ref{fig:PIV_images_exp}a-b. The two images depicted are selected such that the second image is separated from the first by a finite number of propeller revolutions. In the ideal case where the propeller maintains a perfectly constant rotational speed, the two images would be identical, capturing the same geometric phase of the blade. However, since the RPM of the propeller is not perfectly stable, the two images differ and correspond to two distinct phases. This difference is emphasized in Figure \ref{fig:PIV_images_exp}c, which represents the quantity
\begin{equation}
\epsilon_{PP}(x,y)= \frac{[A(x,y)-B(x,y)]^2}{A(x,y)^2}
\label{eq:epsilon}
\end{equation}
where $A$ and $B$ are two different images. In principle, these two images might also differ owing to the variations of the laser pulse intensity in time. Nevertheless, since they correspond to the same laser pulse, this possibility is neglected. \\
Following the same approach explained above, the parameter $k_e$ was set equal to $k_e=7$ and the result of the ISOMAP is shown in Figure \ref{fig:Isomap_exp}.
Since the propeller has two blades, there is a symmetry that reduces the possible phase values between $0$ and $\pi$.
The flow fields obtained from the PIV process can be organized following the sequence illustrated in Figure \ref{fig:Isomap_exp}. Once the flow fields are sorted, phase-averaging can be performed, as detailed in Section \ref{subsec:Phase_Average}. The influence of the number of clusters ($N_{phase}$) on the phase-averaging process is analyzed by evaluating the error $\epsilon_{PP}$, defined in Equation \ref{eq:epsilon}. Here, $A$ and $B$ represent two consecutive images of the flow field within the sorted sequence.
Figure \ref{fig:error}a shows that $\epsilon_{PP}$ decreases as $N_{phase}$ increases, indicating proper sorting of the flow fields. A higher $N_{phase}$ results in greater similarity between consecutive phase-averaged flow fields, further validating the sorting process. This finding is supported by Figure \ref{fig:error}b, which shows the mean standard deviation of the flow fields within the same cluster. The observed reduction in standard deviation with increasing $N_{phase}$ confirms that selecting images that are close in the 2D ISOMAP space corresponds to phase-adjacent snapshots. This outcome aligns with the expectation that phase-adjacent snapshots exhibit similar flow field structures. It is also worth explicitly mentioning that the increase in $N_{phase}$ while decreasing the error in the phase sorting does not automatically enable converged phase averages. On the contrary, an increase in $N_{phase}$ will correspond to a greater number of snapshots required to obtain converged statistics. \\
Thus, it is possible to average the flowfields of adjacent snapshots to compute phase-averaged values.
Increasing the number of phase intervals improves phase resolution, but a sufficient number of images per interval is required to ensure meaningful phase averages.
In Figure \ref{fig:omega}, an example of phase-averaging is reported. Here, $N_{phase}$ was set equal to 24, chosen as a compromise between phase discretization and ensuring enough images per phase cluster for a representative average. To further refine the averaging process, a Gaussian weighting window was applied within each phase cluster, giving more weight to images near the center and progressively less to those at the cluster edges. In this figure, the out-of-plane vorticity $\Omega_z$ normalized with respect to $V_{tip}/D$ (where $V_{tip}$ is the velocity at the propeller's tip and $D$ is the propeller's diameter), is shown for six different phases out of the total 24 employed for the phase averaging process. Specifically, $\theta_i$ with $i=1\dots 6$ correspond to panels \textit{a}, \textit{b}, \textit{c}, \textit{d}, \textit{e}, and \textit{f}, respectively. The flow evolves from bottom to top. The sequence confirms that the phase sorting process is indeed performing as expected, showing a clear streamwise evolution of the tip vortices generated by the propeller's blades. 
To further validate the accuracy of the phase ordering, a vortex tracking approach was applied across the sequence of phase-ordered flowfields. Specifically, the vorticity field was set to isolate the vortex regions, identifying the areas where $\Omega_z > (\Omega_z)_{thr}$. For each detected vortex, the centroid was computed, and the one corresponding to the lowest vortex was selected. This process was repeated for all phases, ensuring that each centroid was associated with the closest one from the previous phase, effectively tracking the vortex evolution. The tracked vortex positions are marked with a green dot in Figure \ref{fig:omega}. As shown in the figure, the vortex moves consistently in the axial direction, in agreement with the expected physical behavior dictated by the blade’s rotation.  

\section{Conclusions}
This work introduces and demonstrates the capabilities enabled by an \emph{a posteriori} data-driven framework for phase-averaged analyses of periodic phenomena and, more in the specific, of optically-measured flows. The approach leverages the ISOMAP algorithm to extract phase information from randomly acquired, uncorrelated snapshots. The methodology enables accurate reordering of the captured data in the phase space and facilitates the computation of phase-averaged flow fields without relying on synchronized data acquisition systems. The proposed framework has been validated through synthetic images and using the experimental datasets of a twin-propeller setup. In both cases, the algorithm shows the capability to robustly capture the periodic characteristics of flow fields with minimal assumptions on acquisition constraints.

The results highlight the framework's potential to overcome challenges associated with traditional phase-averaging methods, such as the strict requirement of precise synchronization in experimental setups. This is particularly advantageous for studying propeller-induced wake flows or other periodic phenomena where synchronization is impractical or cost-prohibitive. Additionally, the capability of applying this technique to archival data makes it a flexible and resource-efficient tool in experimental fluid dynamics.

Future extensions could focus on optimizing the dimensionality reduction step for higher-dimensional datasets or incorporating the framework into real-time processing pipelines. Furthermore, its application could be extended to scenarios involving multi-phase flows or high-Reynolds-number turbulent conditions, where phase-resolved analysis is critical for understanding the flow dynamics. This framework thus establishes a scalable and accessible foundation for advancing phase-averaged methodologies in the study of complex fluid systems.

\section*{Acknowledgments}
This study was carried out within the MOST – Sustainable Mobility National Research Center and received funding from the European Union Next-GenerationEU (PIANO NAZIONALE DI RIPRESA E RESILIENZA (PNRR) – MISSIONE 4 COMPONENTE 2, INVESTIMENTO 1.4 – D.D. 1033 17/06/2022, CN00000023). This manuscript reflects only the authors’ views and opinions, neither the European Union nor the European Commission can be considered responsible for them. SM was supported by a scholarship funded by the Scalability Grant "Winded". 
\bibliographystyle{plainnat} 
\bibliography{bib_file}

\end{document}